\documentclass[conference,10pt,letterpaper]{IEEEtran}
\usepackage{cite}
\usepackage{amsmath,amssymb,amsfonts}
\usepackage{algorithmic}
\usepackage{graphicx}
\usepackage{textcomp}
\usepackage{xcolor}
\usepackage{multicol}
\usepackage{supertabular}
\def\BibTeX{{\rm B\kern-.05em{\sc i\kern-.025em b}\kern-.08em
    T\kern-.1667em\lower.7ex\hbox{E}\kern-.125emX}}

\renewcommand{\thesection}{\textbf{\Roman{section}}}

\usepackage{hyperref}
\begin{document}

\title{\fontsize{24}{24}\selectfont{FerroHEMTs: High-Current and High-Speed All-Epitaxial AlScN/GaN Ferroelectric Transistors}}
\author{\fontsize{11}{11}J. Casamento\textsuperscript{1}$^\dagger$, K. Nomoto\textsuperscript{2}$^\dagger$, T. S. Nguyen\textsuperscript{1}, H. Lee\textsuperscript{2}, C. Savant\textsuperscript{1}, L. Li\textsuperscript{2}, A. Hickman\textsuperscript{2}, T. Maeda\textsuperscript{3},\\ J. Encomendero\textsuperscript{2}, V. Gund\textsuperscript{2}, 
A. Lal\textsuperscript{2}, J. C. M. Hwang\textsuperscript{1,2}, H. G. Xing\textsuperscript{1,2,3}, and D. Jena\textsuperscript{1,2,3}\\
\fontsize{10}{12}\selectfont \textsuperscript{1}Department of Materials Science and Engineering, Cornell University, Ithaca, NY 14853, USA, \\ \textsuperscript{2}School of Electrical and Computer Engineering, Cornell University, Ithaca, NY 14853, USA\\\textsuperscript{3}Kavli Institute at Cornell for Nanoscale Science, Cornell University, Ithaca, NY 14853, USA\\$^\dagger$Equal Contribution. Email: jac694@cornell.edu, kn383@cornell.edu
}

\maketitle
\noindent \textbf{\textit{Abstract}}---We report the first observation of ferroelectric gating in AlScN barrier wide-bandgap nitride transistors.  These FerroHEMT devices realized by direct epitaxial growth represent a new class of ferroelectric transistors in which the semiconductor is itself polar, and the crystalline ferroelectric barrier is lattice-matched to the substrate.  The FerroHEMTs reported here use the thinnest nitride high-K and ferroelectric barriers to date to deliver the highest on-currents at 4 A/mm, and highest speed AlScN transistors with $f_{MAX}>$ 150 GHz observed in any ferroelectric transistor.  The FerroHEMTs exhibit hysteretic $I_d - V_{gs}$ loops with subthreshold slopes below the Boltzmann limit. A control AlN barrier HEMT exhibits neither hysteretic, nor sub-Boltzmann behavior.  While these results introduce the first epitaxial high-K and ferroelectric barrier technology to RF and mm-wave electronics, they are also of interest as a new material platform for combining memory and logic functionalities in digital electronics.

\section{\textbf{Introduction}}

The report of ferroelectric sputtered AlScN in 2019 \cite{jap19_fichtner_alscn_ferroelectric} indicated the tantalizing possibility of introducing ferroelectric barriers into GaN RF and mm-wave transistors. One can pursue this either by integrating sputtered layers on epitaxial channels, or by direct epitaxy of AlScN on GaN transistors \cite{apl17_nrl_mbe_scaln_hemt}.  Direct epitaxial AlScN on GaN revealed high piezoelectric coefficients \cite{apl20_joe_scaln_mbe}. When the leakage was reduced with increased Sc source purity \cite{aplMat21_joe_ScAlN_multilayer_purity}, epitaxial hi-K dielectric constants up to 20 were discovered \cite{apl22_casamento_AlScN_hi_K}.  These epitaxial AlScN layers exhibited a low coercive field of $\sim$ 1 MV/cm \cite{arxiv21_joe_alscn_ferro_lowEc} compared to those of non-epitaxial sputtered layers reported in \cite{jap19_fichtner_alscn_ferroelectric}.  Other reports by epitaxy showed higher coercive fields \cite{apl21_michigan_mbe_scaln_ferroelectric}, and memory functionality \cite{aelm22_umich_ferro_nitride_memory}.  The RF and mm-wave properties of AlScN barrier HEMTs have been studied \cite{kazior2019,green2019,cheng2021}, but ferroelectric transistor behavior has not been reported.  

Here we report ferroelectric gating behavior of epitaxial AlScN barrier HEMTs and contrast its ultra high-current, high-speed, and sub-Boltzmann performance to a control HEMT with a non-ferroelectric AlN barrier.  The measured FerroHEMT behavior is consistent with a low coercive field of epitaxial AlScN.  All measurements presented in this work were performed at room temperature.

\section{\textbf{Epitaxial Growth}}

\begin{figure*}
\centering
  \includegraphics[width=0.58\linewidth]{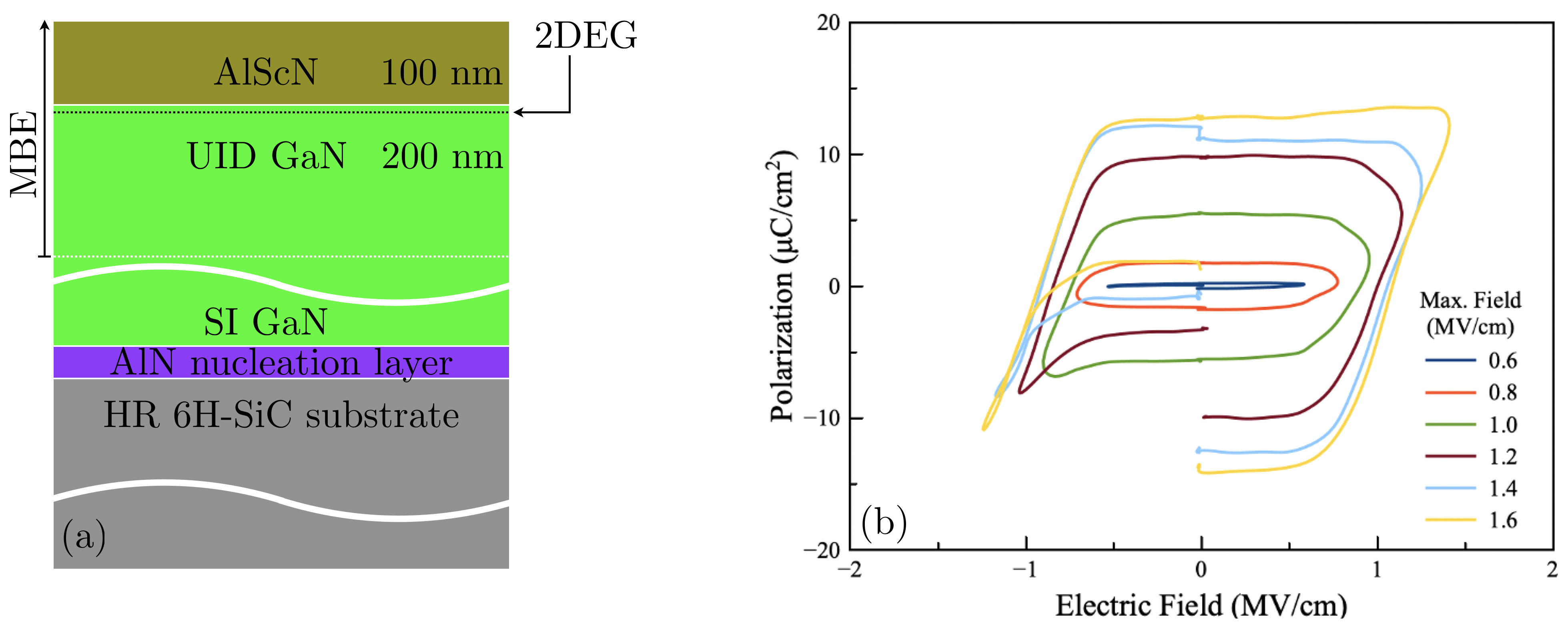}
  \caption{(a) Epitaxial AlScN/GaN heterostructure indicating a polarization-induced 2DEG at the heterojunction. (b) Measured $P-E$ loops on diodes with top electrode and the 2DEG of (a) as the bottom electrode.  Ferroelectric loops are observed for the heterostructure.}
  \label{fig1}
\end{figure*}

Fig. \ref{fig1}(a) shows a test layer structure comprising of a 200 nm unintentionally doped (UID) GaN layer grown by MBE followed by a 100 nm AlScN layer with 18\% Sc. A polarization-induced 2D electron gas (2DEG) of density $1.8 \times 10^{13}$/cm$^2$ and mobility of $377$ cm$^2$/V$\cdot$s was observed in this heterostructure by Hall-effect measurement at room temperature.  Using a deposited top electrode and the 2DEG as a bottom electrode, and applying positive-up-negative-down (PUND) voltage pulses and tracking the currents, the polarization-electric field (P-E) loops extracted are shown in Fig. \ref{fig1}(b) which indicate the epitaxial AlScN layer on UID GaN is ferroelectric with a coercive field $E_c \sim 0.9$ MV/cm.    

A control AlN/GaN HEMT and a 14\% targeted Sc composition AlScN/AlN/GaN FerroHEMT structure were grown by MBE directly on semi-insulating 6H-SiC substrates with a 300 nm AlN nucleation layer and a 1 $\mu$m GaN buffer layer using methods reported in \cite{apl22_casamento_AlScN_hi_K} to study ferroelectric gating behavior.  Figs. \ref{fig2}(a) and (c) show the corresponding energy band diagrams from the top surface, calculated using self-consistent Schr\"{o}dinger-Poisson solutions. 2DEG channels are expected at the heterojunctions as shown as $\psi_{1}^2$ at respective depths. To form Ohmic contacts to these 2DEGs, lithographically defined etching and regrowth of heavily doped n+ GaN ($N_d \sim$ 10$^{20}$/cm$^3$ Si) was performed by MBE.  Figs. \ref{fig2}(b) and (d) show the measured 2DEG densities, mobilities, and sheet resistances over various dies of the wafer measured by room-temperature Hall effect.  The 2DEG densities are consistent with the expected values from the calculation. 

\section{\textbf{Device design and fabrication}}

Figs. \ref{fig3}(a) and (b) show the resistances of metal-regrown n$^+$GaN (low) and n$^+$GaN-2DEG (moderate, not exceptionally low).  Figs. \ref{fig4}(a) and (b) show the control AlN barrier HEMT and the AlScN barrier FerroHEMT cross sections respectively. Figs. \ref{fig4}(c) shows the representative device process flow of HEMTs with regrown contacts. A SiO$_2$/Cr hard mask defined the source/drain regions. Ti/Au source/drain was deposited on the n$^+$GaN, and Ni/Au gate was deposited. Electron beam lithography (EBL) process was performed to fabricate T-gate devices for RF and mm-wave performance. Figs. \ref{fig4}(d) and the inset show SEM images of the final transistor structures.  Gate lengths ranged from 90 nm to 18 $\mu$m. 

\section{\textbf{Results and discussion}}

Figs. \ref{fig5}(a) and (b) show the measured $I_d - V_{ds}$ output characteristics and $I_{d}-V_{gs}$ transfer characteristics of the control AlN/GaN HEMT sample respectively. An on current of $\sim 1.5 $ A/mm with an on resistance of $R_{on}=1.34$ $\Omega \cdot$mm, an on/off ratio of $10^{6}$ limited by gate leakage, with a threshold voltage of $\sim -3.5$ V were observed. Fig. \ref{fig5}(c) shows a peak transconductance of $\sim 0.5$ S/mm with a barrier thickness of 2 nm GaN + 2 nm AlN.  The subthreshold slope shown in Fig. \ref{fig5}(d) indicates very good normal transistor behavior, grazing the ideal Boltzmann limit of $\sim 60$ mV/decade.  The control HEMT thus exhibits excellent performance, as borne out by its RF performance discussed subsequently.

Fig. \ref{fig6} shows the transistor characteristics when a 5 nm thick epitaxial AlScN barrier layer is added between the AlN and the GaN cap layer, as indicated in Figs. \ref{fig4}(a)-(b). From Fig. \ref{fig6}(a), the maximum $I_d$ still reaches $\sim 1.5$ A/mm at the same gate voltage, {\em in spite of a more than double barrier thickness} of 2 nm GaN + 5 nm AlScN + 2 nm AlN compared to the control sample.  This is a result of the high-K dielectric property of AlScN as was reported in \cite{apl22_casamento_AlScN_hi_K}. The $I_d - V_{ds}$ curves indicate a higher output conductance, but a far larger difference from the control sample is observed in the transfer characteristics in Fig. \ref{fig6}(b). A counterclockwise (CCW) hysteresis loop develops in the subthreshold characteristics. A  sub-Boltzmann steep slope of 23.6 mV/decade is observed for the on$\rightarrow$off voltage sweep.  Such repeatable loops are observed in multiple devices.  This translates to a hysteretic transconductance curve as seen in Fig. \ref{fig6}(c) and drain current as seen in Fig. \ref{fig6}(d).  Note that though the peak transconductance is lower than the control HEMT, the hi-K AlScN helps maintain a high value in spite of a more than double barrier thickness.

The hysteresis window of the threshold voltage is between $1.0 - 2.0$ V as seen in Figs. \ref{fig6}(c) and (d).  These observations are strong signatures of both high-K dielectric, and ferroelectric gating behavior.  Based on the AlScN barrier thickness, a voltage drop $E_{c} \times t_{AlScN} \sim 0.5 $ V across the AlScN layer is consistent with a low $E_c \sim 1.0$ MV/cm. A large portion of the gate voltage still drops across the GaN and AlN layers between the gate metal and the 2DEG channel.  

Fig. \ref{fig7}(a) shows the hysteresis loop measured on Die 7 of Fig. 2 (d).  Several CCW hysteresis loops based on the voltage step of the measurement are shown in Fig. \ref{fig7}(a), and the corresponding substhreshold slopes are plotted in Figs. \ref{fig7}(b) and (c).  While the majority of the sub-Boltzmann steep slopes are observed in the leftgoing voltage sweeps, some also appear in the rightgoing sweeps.  It is well known that trapping behavior could lead to false conclusions of ferroelectricity. The CCW loops for n-channels is a strong evidence of ferroelectric FETs. Moreover, the absence of such behavior in the control HEMT sample, and the symmetry, voltage width, and sub-Boltzmann slopes of the FerroHEMT conclusively indicates ferroelectric gating in the all-epitaxial AlScN/GaN devices.

Fig. \ref{fig8}(a) shows the high-frequency characteristics of the AlN barrier control HEMT.  It exhibits excellent cutoff frequencies of $f_{T}/f_{MAX} = 126/304$ GHz for a gate length of $L_g = 90$ nm.  The device dimensions and bias conditions are indicated in the plot. Fig. \ref{fig8}(b) shows the measured values for the FerroHEMT of similar dimensions but different bias conditions due to the shifted threshold characteristics.  It exhibits lower cutoff frequencies of $f_{T}/f_{MAX} = 78/156$ GHz.  Even though the values of the AlScN FerroHEMT are lower than the control AlN barrier HEMT, they are the highest reported to date for AlScN barrier transistors, as indicated in Fig. \ref{fig8}(c). Morever, they are the fastest {\em ferroelectric} transistors reported to date: a maximum FerroHEMT $f_{MAX} = 168$ GHz is measured.  The higher speed of the control sample is due to the (expected) higher maximum $g_{m,ext} = 0.616$ S/mm of the control HEMT compared to $g_{m,ext} = 0.475$ S/mm for the FerroHEMT, indicating higher speed FerroHEMTs are possible with further scaling.  Fig. \ref{fig8}(d) shows the scaling of the output current $I_{d}^{max}$ in the AlScN FerroHEMTs when $L_g$ is scaled from 18 $\mu$m to 90 nm.  An exceptionally high value of 2.5 A/mm is observed for an {\em optical} gate length of 1 $\mu$m.  The deep submicron EBL gates reach record values of 4 A/mm at 90 nm gate length.  These record high on-currents are enabled by the high 2DEG density generated by the large difference in polarization between GaN and the epitaxial AlScN layers.

\section{\textbf{Conclusions and future work}}

\begin{figure*}
\centering
  \includegraphics[width=1.0\linewidth]{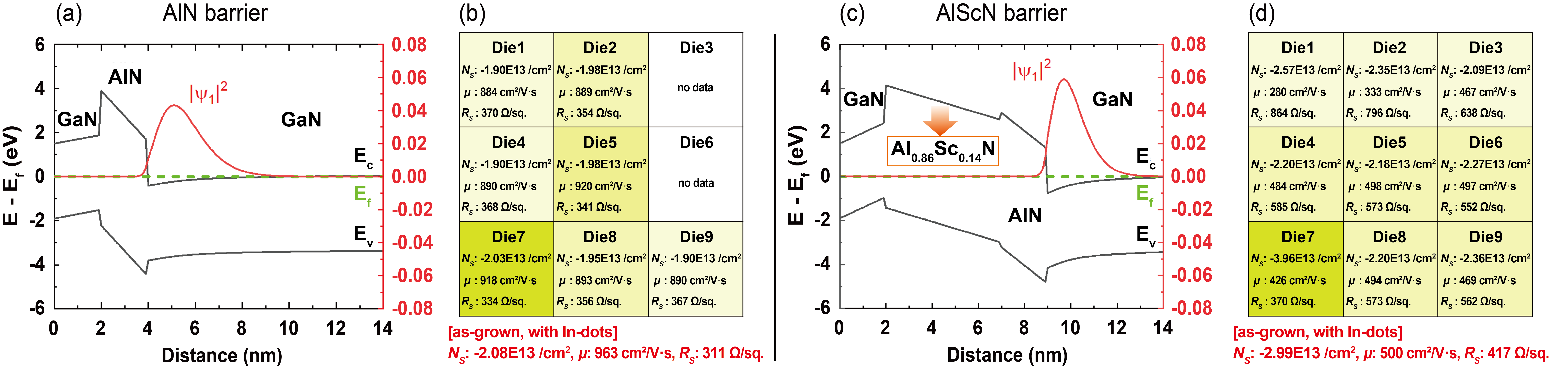}
  \caption{(a) Energy band diagram of control HEMT. (b) Hall-effect data of control HEMT. (c) Energy band diagram of FerroHEMT. (d) Hall-effect data of FerroHEMT.  Though there are some non-uniformities, the measured 2DEG densities are consistent with the calculations.}
  \label{fig2}
\end{figure*}

\begin{figure*}
\centering
  \includegraphics[width=1.0\linewidth]{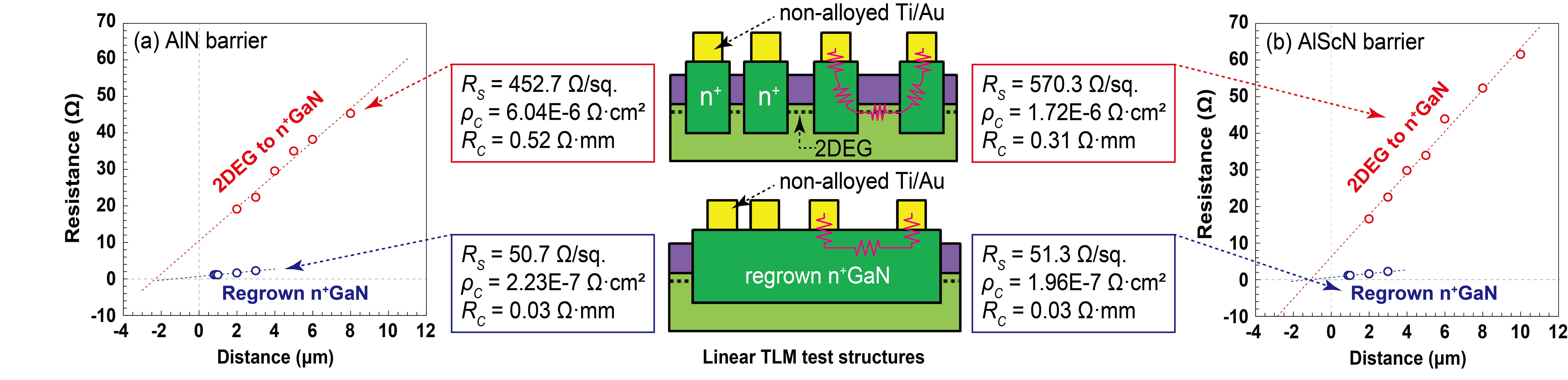}
  \caption{(a) Contact resistances for the control HEMT between the metal and the regrown n$^+$GaN regions and the n$^+$GaN regions and 2DEGs. (b) Corresponding contact resistances for the FerroHEMT structure. }
  \label{fig3}
\end{figure*}

\begin{figure*}
\centering
  \includegraphics[width=1.0\linewidth]{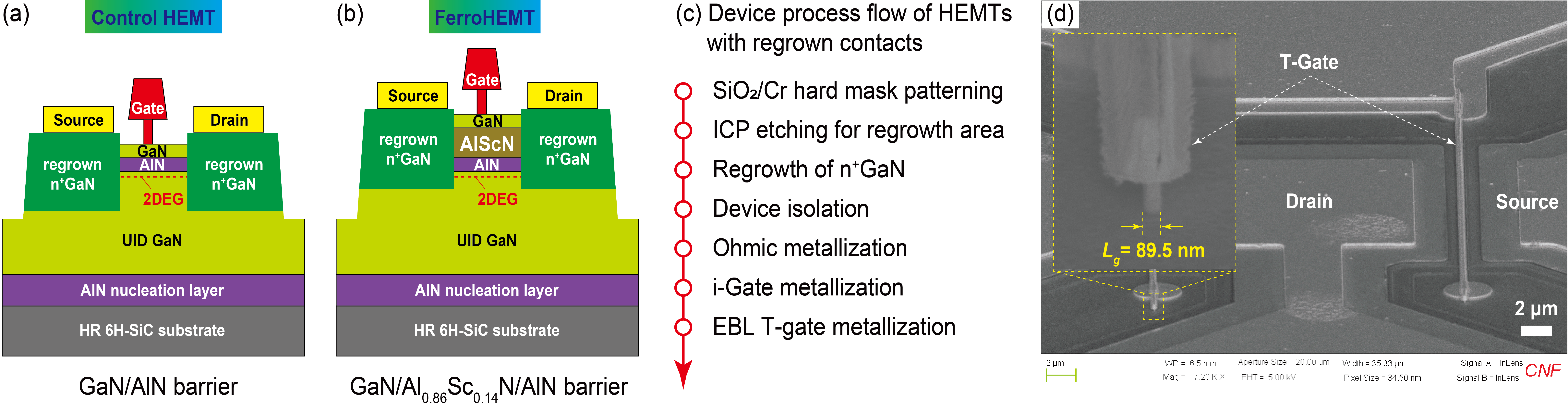}
  \caption{(a) Control HEMT cross section schematic showing the regrown contacts, gate, and channel regions. (b) Corresponding schematic for the FerroHEMT has an extra 5 nm AlScN layer in the barrier layer. (c) The representative device process flow of III-Nitride HEMTs with regrown contacts. (d) SEM image of a processed HEMT, and inset shows Zoomed-in image of a T-gate.}
  \label{fig4}
\end{figure*}

\begin{figure*}
\centering
  \includegraphics[width=0.95\linewidth]{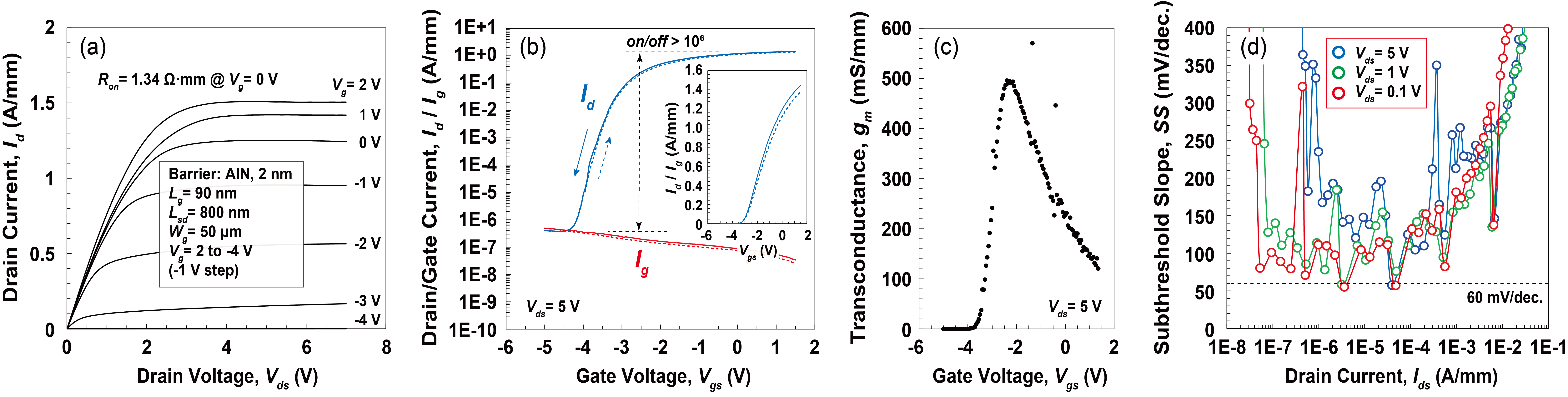}
  \caption{Measured characteristics of the control HEMT with an AlN barrier: (a) Output characteristics, (b) Transfer characteristics in the log and (inset) linear scale, (c) Transconductance vs. $V_{gs}$, and (d) subthreshold slope vs. $I_d$. }
  \label{fig5}
\end{figure*}

\begin{figure*}
\centering
  \includegraphics[width=0.9\linewidth]{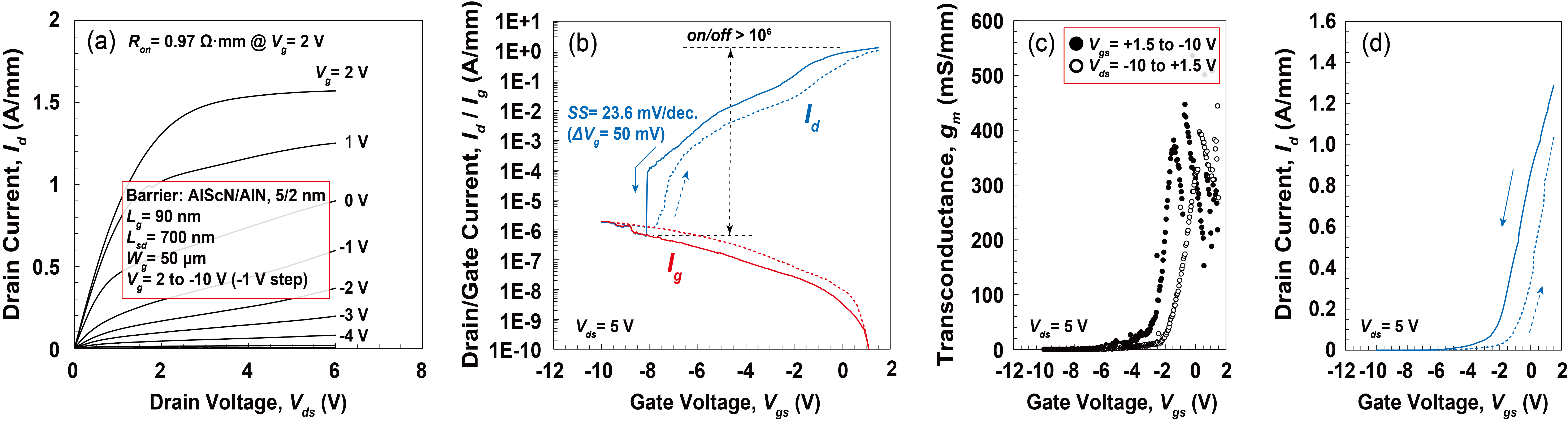}
  \caption{Measured characteristics of the FerroHEMT with an AlScN barrier: (a) Output characteristics, (b) Transfer characteristics in the log scale showing a counterclockwise hysteresis and sub-Boltzmann slope, (c) Transconductance vs. $V_{gs}$ showing hysteresis, and (d) Linear $I_d$ vs $V_{gs}$ showing hysteresis.}
  \label{fig6}
\end{figure*}

\begin{figure*}
\centering
  \includegraphics[width=0.83\linewidth]{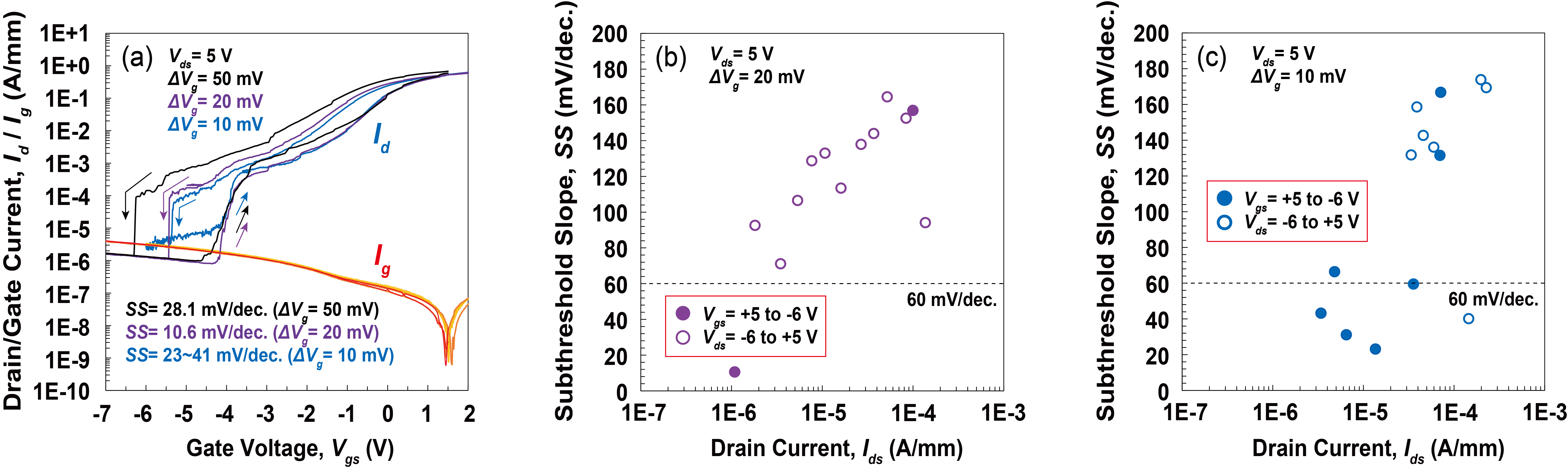}
  \caption{(a) Subthreshold characteristics of a FerroHEMT showing counterclockwise hysteresis loops and steep slopes for various bias steps.  Sub-Boltzmann behavior observed in the subthreshold slopes for (b) 20 mV gate sweep steps and (c) 10 mV gate sweep steps. }
  \label{fig7}
\end{figure*}

\begin{figure*}
\centering
  \includegraphics[width=0.95\linewidth]{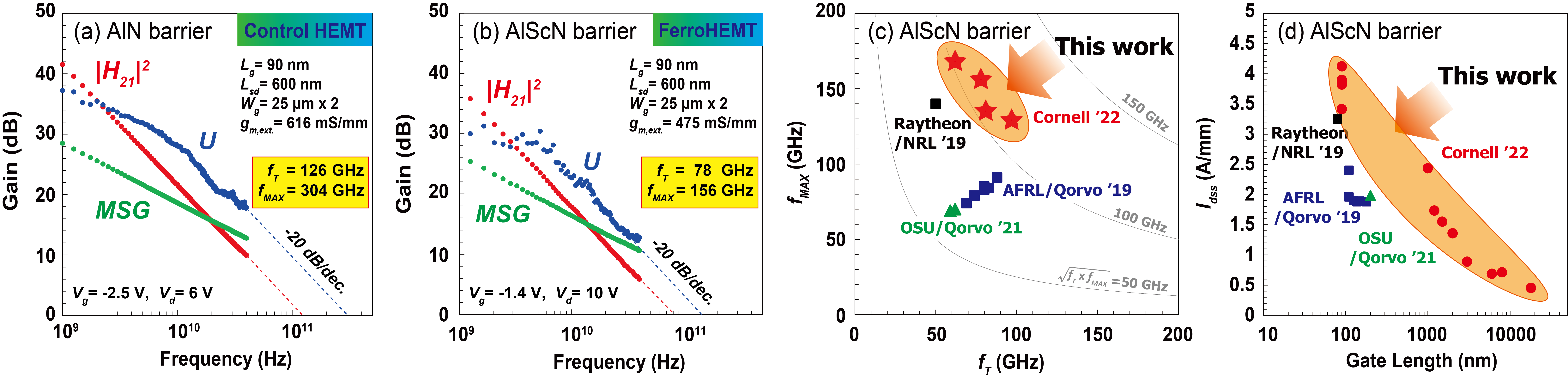}
  \caption{Measured cutoff frequencies at $L_{g}=90$ nm of (a) the Control HEMT, and (b) the AlScN barrier FerroHEMT.  (c) Speed benchmarks of AlScN HEMTs with previous reports~\cite{kazior2019,green2019,cheng2021}. (d) Scaling of output current with gate length.  In addition to being the fastest AlScN HEMTs, the FerroHEMT data reported here also represents the highest speed and highest on-current ferroelectric transistors realized to date.}
  \label{fig8}
\end{figure*}

The moderate FerroHEMT channel mobility in Fig. \ref{fig2}(d) can be improved by 3$\times$ for better RF and mm-wave performance.  The high-K value of AlScN will increase the breakdown voltage in FerroHEMTs by reducing the gate leakage \cite{apl22_casamento_AlScN_hi_K}.  A negative DIBL effect is predicted \cite{edl19_sayeef_ferro_negative_dibl}, but this will be achievable with careful geometrical design of future epitaxial FerroHEMTs.  Thus FerroHEMTs with thinnest epitaxial ferroelectric barriers show steep subthreshold slopes, and deliver the highest on-currents (4 A/mm), and highest speed ($f_{MAX}>$ 150 GHz) in all ferroelectric transistors.  They present a new class of transistors with the potential to blur the boundaries between memory, logic, and communication devices.

\section*{Acknowledgment}
This work was supported in part by the SRC and DARPA through the Joint University Microelectronics Program (JUMP), by the DARPA TUFEN program, and performed at the Cornell NanoScale Facility, an NNCI member supported by NSF Grant No. NNCI-2025233.

% Dummy (non-printed) citations to trigger automatic sorting and inclusion of the references.

\nocite{fontnote}
\nocite{fuller}
\nocite{vidmar}
\nocite{clarke}
\nocite{reber}
\nocite{yorozu}
\nocite{alqueres}

% INSERT REFERENCE LIST
\bibliography{references}

\end{document}